\documentclass[
]{ceurart}

\usepackage{caption}
\usepackage{listings}

\begin{document}

\copyrightyear{2025}
\copyrightclause{Copyright for this paper by its authors. Use permitted under Creative Commons License Attribution 4.0 International (CC BY 4.0).}

\conference{ISWC 2024 Special Session on Harmonising Generative AI and Semantic Web Technologies, November 13, 2024, Baltimore, Maryland}

\title{LLM-based SPARQL Query Generation from Natural Language over Federated Knowledge Graphs}

\author[1]{Vincent Emonet}[%
orcid=0000-0002-1501-1082,
email=vincent.emonet@sib.swiss,
]
\cormark[1]
\address[1]{SIB Swiss Institute of Bioinformatics, Switzerland}

\author[1]{Jerven Bolleman}[%
orcid=0000-0002-7449-1266,
email=Jerven.Bolleman@sib.swiss,
]

\author[1]{Severine Duvaud}[%
orcid=0000-0001-7892-9678,
email=severine.duvaud@sib.swiss,
]

\author[1]{Tarcisio {Mendes de Farias}}[%
orcid=0000-0002-3175-5372,
email=tarcisio.mendes@sib.swiss,
]

\author[1]{Ana Claudia Sima}[%
orcid=0000-0003-3213-4495,
email=ana-claudia.sima@sib.swiss,
]

\cortext[1]{Corresponding author.}

\begin{abstract}
We introduce a Retrieval-Augmented Generation (RAG) system for translating user questions into accurate federated SPARQL queries over bioinformatics knowledge graphs (KGs) leveraging Large Language Models (LLMs). To enhance accuracy and reduce hallucinations in query generation, our system utilises metadata from the KGs, including query examples and schema information, and incorporates a validation step to correct generated queries. The system is available online at \href{https://chat.expasy.org}{chat.expasy.org}. 
\end{abstract}



\maketitle

\section{Introduction}

In bioinformatics, the ability to query complex knowledge graphs (KGs) is critical for extracting meaningful insights. However, manually crafting SPARQL queries, especially federated queries spanning across multiple connected KGs, can be a time-consuming and challenging task, even for experts. This has led to a growing demand for Knowledge Graph Question Answering (KGQA) systems that can translate natural language queries into SPARQL, bridging the gap between users' questions and the structured data available. 

Large Language Models (LLMs) present an exciting opportunity to address this challenge, offering the potential to automatically generate accurate SPARQL queries from natural language inputs. However, while LLMs have demonstrated impressive capabilities in this area \cite{sima2023potential}\cite{meyer2024assessingsparqlcapabilitieslarge}, current systems struggle to handle large-scale, evolving KGs, such as those in the SIB Swiss Institute of Bioinformatics catalog \cite{sib2024sib}.

In this work, we present a solution designed to assist users of SIB’s bioinformatics KGs\cite{sima2019enabling}, such as UniProt \cite{uniprot2023uniprot}, Bgee \cite{bastian2021bgee} or OMA \cite{altenhoff2024oma}, to explore and query the data available. Our approach leverages LLMs and endpoints metadata to generate SPARQL queries while addressing the challenge of dynamically integrating evolving datasets, without requiring constant retraining. By offering a scalable system\footnote{\href{https://github.com/sib-swiss/sparql-llm}{github.com/sib-swiss/sparql-llm}} that adapts to the complex and changing landscape of bioinformatics knowledge, we aim to significantly reduce the time and expertise needed to query across federated KGs.

\section{System Design}

\begin{figure*}[h]
  \centering
  \includegraphics[width=0.95\textwidth]{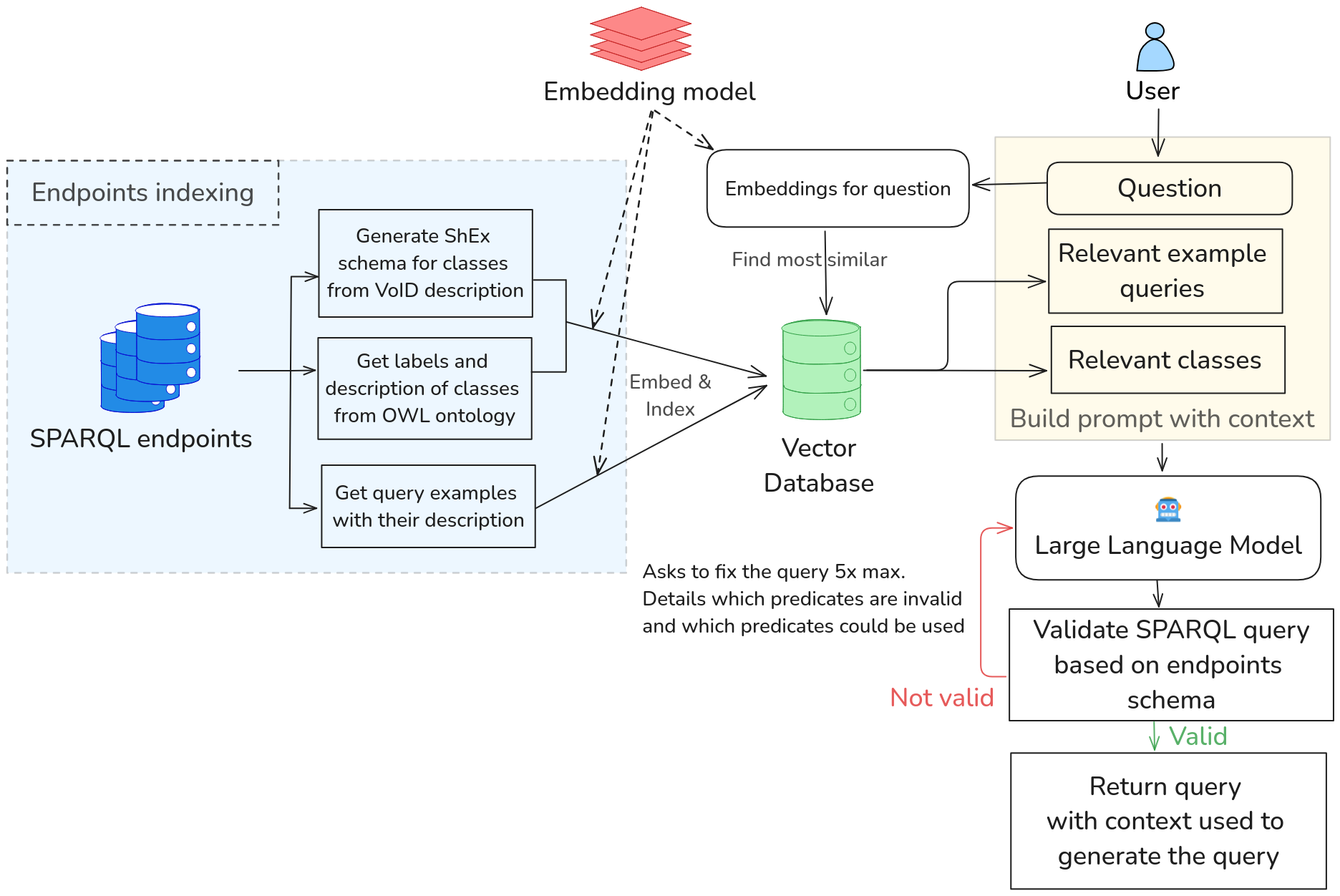}
  \caption{LLM-based SPARQL Query Generator System Architecture.}
  \label{fig:sysarch}
\end{figure*}

The proposed system is illustrated in Figure~\ref{fig:sysarch}. The system takes a list of SPARQL endpoint URLs as input, where each endpoint is expected to include minimal standardized metadata (example queries and VoID\footnote{Vocabulary of Interlinked Datasets \href{https://www.w3.org/TR/void/}{w3.org/TR/void}} descriptions) that can be automatically retrieved and indexed upon initial deployment. We provide an online webpage\footnote{\href{https://sib-swiss.github.io/sparql-editor/check}{sib-swiss.github.io/sparql-editor/check}} allowing to check if a given endpoint contains the required metadata.

The overall data flow of the system can be summarised as: 1) getting the relevant context using embeddings-based similarity search; 2) prompt building leveraging the retrieved context; 3) validating and correcting the query using endpoints schema; 4) presenting query and relevant context to the user.






\subsection{Generating embeddings and indexing}

Example question/query pairs are automatically retrieved from each endpoint via a SPARQL query upon initial deployment. Embeddings are generated for each question and indexed in a vector database, allowing to match user provided inputs based on similarity search. A documented repository \cite{sparqlexamples2024} of example question/query pairs\footnote{\href{https://github.com/sib-swiss/sparql-examples}{github.com/sib-swiss/sparql-examples}} and a CLI tool\footnote{\href{https://github.com/sib-swiss/sparql-examples-utils}{github.com/sib-swiss/sparql-examples-utils}} are provided to help endpoints maintainers define and validate SPARQL query examples for their endpoints in a standardized format.

Additionally, we generate and index shapes for the classes present in each endpoint. For this, we first retrieve the VoID descriptions from each endpoint, which detail the relationships between subject classes and object classes or datatypes via specific predicates. For example, the \textit{Protein} class is linked to the \textit{Gene} class through the \textit{encodedBy} predicate. The VoID generator is available open source\footnote{\href{https://github.com/JervenBolleman/void-generator}{github.com/JervenBolleman/void-generator}} and can be used to generate statistics over any SPARQL endpoint. This information allows us to generate simple, human-readable Shape Expressions (ShEx) for each class. Each object property references a list of classes rather than another shape, making each shape self-contained and interpretable on its own.

We chose to use the ShEx schema language instead of the Web Ontology Language (OWL) for several reasons. First, endpoints may not implement all properties and classes described in the OWL ontologies they utilize, as ontologies operate under the open-world assumption. As a consequence, a more precise schema, based on the actual contents of the endpoint, must be generated for improved accuracy. Additionally, ShEx is specifically designed to express schema constraints, and its syntax is closer to the SPARQL syntax, making it more suitable for defining the structure of data in our context. In contrast, OWL requires separate classes to define properties with specified domains and ranges, resulting in a more complex and verbose representation. As a side effect, the increased verbosity would also incur higher costs due to the increase in number of tokens.

The generated shapes are well-suited for use with a LLM, as they provide information about which predicates are available for a class, and the corresponding classes or datatypes those predicates point to. For example here is the shape generated for a \textit{Disease Annotation}:

\begin{verbatim}
up:Disease_Annotation {
  a [ up:Disease_Annotation ] ;
  up:sequence [ up:Chain_Annotation up:Modified_Sequence ];
  rdfs:comment xsd:string ; up:disease IRI }
\end{verbatim}

Once the shapes have been generated, labels and descriptions are retrieved from the class OWL ontology by querying the endpoint, and then indexed in the vector database for each class. If no labels or descriptions are found, the class URI is used as the label.


General information about the contents of each endpoint, retrieved from the schema.org metadata available at each SPARQL endpoint homepage, are also added to the vector database. This allows to provide general information about the endpoint in case questions about it are asked by the users. 

\subsection{Prompt building}


When a user asks a question, embeddings are generated for the question, and the 20 most similar questions and 15 closest class labels are retrieved from the vector database. A limit is used rather than a similarity threshold to ensure the model has access to enough examples to determine whether the question can be answered based on the available endpoint data. Moreover, even if the retrieved examples have a low similarity score, they still provide hints on how to write the query, and can be useful when combined with the classes schemas. The retrieved questions and their associated queries, as well as the classes label and their associated schema, are added to the prompt alongside the user question. A concrete example of a generated prompt can be found online \footnote{\href{https://github.com/sib-swiss/sparql-llm/blob/main/notebooks/EXAMPLE_PROMPT.md}{github.com/sib-swiss/sparql-llm/blob/main/notebooks/EXAMPLE\_PROMPT.md}}.




\subsection{Validating generated queries}

To mitigate potential errors (e.g., hallucinations), we developed a method to validate federated SPARQL queries based on the VoID description of the endpoints. First, the validator parses the SPARQL query generated, which can be a federated query. Next, the validator extracts triple patterns, identifies the endpoint(s) where these will be executed, and whether the triple patterns comply with the expected schema. For each subject that has an explicit class assigned, the system will check if the stated predicates are available for this class according to the precomputed ShEx schema. The system will also try to infer the classes of other connected entities that do not have a class explicitly defined.




The validator produces a list of human-readable errors describing which predicates are incorrect, the associated subject or class, and a list of valid predicates based on the schema. The errors and  corrections are then provided back to the LLM to help correct the wrong query, for example: "\textit{Subject ?disease with type up:Disease in endpoint https://sparql.uniprot.org/sparql does not support the predicate rdfs:label. It can have the following predicates: skos:altLabel, rdfs:comment, up:mnemonic, skos:prefLabel, rdfs:seeAlso}".






\subsection{Context transparency and feedback}

The "\textit{See relevant references}" button in the chat interface allows users to browse the retrieved queries and classes provided as context to the LLM, along with their similarity scores. This feature enables users to investigate references and understand the reasons behind any incorrect generated query.


All user questions are stored in a log file. Additionally, the chat interface includes a simple user feedback mechanism through like/dislike buttons. When one of these buttons is clicked, the entire conversation is stored in a file on the server. A notebook is provided to display all logged conversations in a human-readable way to help maintainers analyze feedbacks.

\section{Implementation}

The different components of the system are modular, published as a python package\footnote{\href{https://pypi.org/project/sparql-llm}{pypi.org/project/sparql-llm}}, and can be reused in other systems. The main components included are: 1) A module to automatically generate a human-readable ShEx schema for each class extracted from the VoID description of an endpoint, and generate documents to load in a vector database, compatible with the LangChain framework; 2) A module to automatically retrieve query examples from an endpoint and generate documents to load in a vector database; 3) A module to automatically validate a federated SPARQL query based on each endpoint VoID description, providing human-readable error messages and suggested corrections.



The system can easily support any LLM that is served via an OpenAI-compatible API. As of today, a lot of open-source inference tools and LLM cloud providers use this standard. While we primarily rely on OpenAI models, we have also been testing the system with models such as LLaMA and Mixtral.

The fastembed\footnote{\href{https://qdrant.github.io/fastembed/}{qdrant.github.io/fastembed}} library is used with the model \textit{BAAI/bge-large-en-v1.5} to generate text embeddings, chosen for its speed and high performance on the Massive Text Embedding Benchmark  \cite{muennighoff2022mteb}\cite{excoffier2024generalist}\footnote{\href{https://huggingface.co/spaces/mteb/leaderboard}{huggingface.co/spaces/mteb/leaderboard}}. The Qdrant\footnote{\href{https://qdrant.tech/}{qdrant.tech}} vector database is used to store the embeddings and perform cosine similarity search.



The system is containerised to facilitate reusability and deployment via docker/podman compose.


\section{Evaluation and discussion}


We designed a preliminary test suite of 13 questions, each paired with a reference query, ensuring none are present in the examples seen by the system. The questions require reasoning over the context and specify clear retrieval tasks. For each question, the system generates a query, which we run and compare against the expected results. To account for LLM variability, each question is tested 3 times.

We assess the system's performance across three configurations: 1) baseline LLM without RAG, 2) RAG without validation, and 3) RAG with validation and correction of the generated query. This approach allows us to evaluate the contribution of each system component in improving query generation accuracy and helps detect any regressions when modifications are introduced. Additionally, the test suite is adaptable for testing different LLMs, enabling a comparison of their effectiveness in handling the questions. The results are summarised in Table~\ref{tab:model_comparison}, price is computed from average per request.

The results indicate that larger LLMs perform significantly better overall, efficiently extrapolating from provided examples. In contrast, query validation is particularly valuable for smaller LLMs, not only improving accuracy but also ensuring the system generates queries that retrieve at least some relevant results.



So far, our primary focus has been on developing a production system to assist users of the SIB SPARQL endpoints in formulating effective queries. In the future, we aim to conduct a more comprehensive evaluation using standardised benchmarks\cite{Dblp-quad}\cite{sciqa2023} and benchmarking frameworks\cite{meyer2024assessingsparqlcapabilitieslarge} to assess the system's precision, usability, and overall performance across a wider range of queries.

We observed that when the system retrieves context that does not fully align with the user’s question, it typically provides an answer that, while imprecise, helps guide the user towards understanding how they might answer the question using available resources. For example, when asked, "What is the function of the esophagus?" the system may return a query that lists anatomical entities with descriptions where "esophagus" appears in the name. Though not directly answering the user’s question, this query still brings the user closer to a solution by suggesting relevant resources. If the retrieved context is insufficiently relevant, the system explicitly states that it cannot answer the question.

Evaluating query results in bioinformatics is particularly challenging, even for human experts, due to the inherent complexity of both the domain-specific questions and the knowledge graph models. These challenges often introduce interpretative flexibility when determining the accuracy or relevance of the system’s outputs. As such, we expect users—typically researchers—to apply a critical approach when assessing the answers provided by the system.

The main weakness we identified in the current system is the tendency to hallucinate entity identifiers that cannot be found in examples, as well as its reliance on inefficient string comparisons that are not well optimized in SPARQL engines. To address these issues, we plan to introduce entity indexing from the endpoints, along with few-shot prompting, to enhance entity extraction and resolution.

\begin{table}
  \captionsetup{format=plain}
  \caption{Test results for different models and RAG approaches across query results categories}
  \label{tab:model_comparison}
  \begin{tabular}{lccccccc}
    \toprule
    Model & Approach & Success & Different Result & No Result & Error & Price (\$)  & F1 \\
    \midrule
    gpt-4o & No RAG & 3 & 0 & 36 & 0 & 0.00478 & 0.08 \\
    gpt-4o & RAG w/o validation & 33 & 0 & 5 & 1 & 0.03707 & 0.85 \\
    gpt-4o & RAG w/ validation & 34 & 3 & 1 & 1 & 0.04781 & 0.91 \\
    \midrule
    gpt-4o-mini & No RAG & 0 & 0 & 9 & 30 & 0.00011 & 0.0 \\
    gpt-4o-mini & RAG w/o validation & 13 & 7 & 18 & 1 & 0.00111 & 0.37 \\
    gpt-4o-mini & RAG w/ validation & 11 & 18 & 9 & 1 & 0.0019 & 0.37 \\
    \midrule
    Mixtral 8x22B & No RAG & 0 & 0 & 16 & 23 & 0.0007 & 0.0 \\
    Mixtral 8x22B & RAG w/o validation & 6 & 11 & 20 & 2 & 0.01073 & 0.18 \\
    Mixtral 8x22B & RAG w/ validation & 10 & 14 & 10 & 5 & 0.02147 & 0.31 \\
    \midrule
    Llama3.1 8B & No RAG & 0 & 0 & 6 & 33 & 8e-05 & 0.0 \\
    Llama3.1 8B & RAG w/o validation & 0 & 0 & 15 & 24 & 0.00144 & 0.0 \\
    Llama3.1 8B & RAG w/ validation & 3 & 2 & 20 & 14 & 0.00405 & 0.08 \\
    \bottomrule
  \end{tabular}
\end{table}










\section{Conclusions}

We have introduced a system for generating federated SPARQL queries leveraging LLMs over real-world KGs, in response to Natural Language questions asked by users. The system currently handles questions over bioinformatics KGs within the SIB Semantic Web of Data, but can easily be reused with other KGs of interest. The system is fully open-source and a demo can be accessed online at \href{https://chat.expasy.org}{chat.expasy.org}.

\begin{acknowledgments}
We acknowledge contributions and fruitful discussions with the SIB Semantic Web Focus Group, as well as funding provided through the Biodata Resources Team at SIB. This work was also supported by the CHIST-ERA grant CHIST-ERA-22-ORD-09 by SNF grant 20CH21\_217482 and the Swiss Open Research Data Grants (CHORD) in Open Science I, a program coordinated by swissuniversities, grant id “Swiss DBGI-KM”. UniProt is supported by the National Eye Institute (NEI), National Human Genome Research Institute (NHGRI), National Heart, Lung, and Blood Institute (NHLBI), National Institute on Aging (NIA), National Institute of Allergy and Infectious Diseases (NIAID), National Institute of Diabetes and Digestive and Kidney Diseases (NIDDK), National Institute of General Medical Sciences (NIGMS), National Institute of Mental Health (NIMH), and National Cancer Institute (NCI) of the National Institutes of Health (NIH) under grant U24HG007822. UniProt activities at the SIB are additionally supported by the Swiss Federal Government through the State Secretariat for Education, Research and Innovation SERI.
\end{acknowledgments}

\bibliography{sample-ceur}

\appendix

\section{Online Resources}

The source code for the Expasy chat system and the reusable modules.

\begin{itemize}
\item Chat system and components source code: \href{https://github.com/sib-swiss/sparql-llm}{github.com/sib-swiss/sparql-llm},
\item
  SPARQL examples example repository: \href{https://github.com/sib-swiss/sparql-examples}{github.com/sib-swiss/sparql-examples},
\item
  CLI to validate SPARQL examples: \href{https://github.com/sib-swiss/sparql-examples-utils}{github.com/sib-swiss/sparql-examples-utils},
\item
  CLI to generate VoID descriptions: \href{https://github.com/JervenBolleman/void-generator}{github.com/JervenBolleman/void-generator},
\item
  Check an endpoint metadata: \href{https://sib-swiss.github.io/sparql-editor/check}{sib-swiss.github.io/sparql-editor/check}.
\end{itemize}

\end{document}